%% file: main.tex
\documentclass[journal,transmag]{IEEEtran}

\usepackage{graphicx}
\usepackage{epstopdf}
\usepackage{epsfig}
\usepackage{subfigure}
\usepackage{stfloats}
\usepackage{url}

\usepackage{times}
\usepackage{amsmath}
\usepackage{amsthm}
\usepackage{amssymb}
\usepackage{cite}
\usepackage{color}

\newtheorem{theorem}{Theorem}

\def\p{{\bf p}}

\def\d{{\bf d}}

\def\x{{\bf x}}
\def\y{{\bf y}}
\def\n{{\bf n}}
\def\z{{\bf z}}

\def\s{{\bf s}}
\def\r{{\bf r}}
\long\def\comment#1{}
\author{\IEEEauthorblockN{Yacov Hel-Or\IEEEauthorrefmark{1}
and Gil Ben-Artzi \IEEEauthorrefmark{2}}

\IEEEauthorblockA{\IEEEauthorrefmark{1} Efi Arazi School of Computer Science,
The Interdisciplinary Center, Herzliya, Israel 30900}
\IEEEauthorblockA{\IEEEauthorrefmark{2} Department of Computer Science, Ariel University, Ariel, Israel 40700}

\thanks{Corresponding author: Y. Hel-Or (email: toky@idc.ac.il).}
}

\title{The Role of Redundant Bases and Shrinkage Functions \\ in Image Denoising}

\begin{document}

\IEEEtitleabstractindextext{%
\begin{abstract}
Wavelet denoising is a classical and effective approach for reducing noise in images and signals. Suggested in 1994 \cite{Donoho94}, this approach is carried out by rectifying the coefficients of a noisy image in the transform domain, using a set of scalar shrinkage function (SFs). A plethora of papers deals with the optimal shape of the SFs and the transform used, where it is known that applying the SFs in redundant bases provides improved results.  This paper provides a complete picture of the interrelations between the transform used, the optimal shrinkage functions, and the domains in which they are optimized. In particular, we show that for subband optimization, where each SF is optimized independently for a particular band, optimizing the SFs in the spatial domain is always better than or equal to optimizing the SFs in the transform domain.  For redundant bases, we provide the expected denoising gain we may achieve, relative to the unitary basis, as a function of the redundancy rate. 
\end{abstract}

% Note that keywords are not normally used for peerreview papers.
\begin{IEEEkeywords}
Wavelet transforms, image restoration, image denoising, shrinkage denoising, cycle spinning, noise removal, overcomplete representation.
\end{IEEEkeywords}}

\maketitle

\input{intro}

\input{redundant}

\input{bounds}

\input{results}

\input{appndx}

\section*{Acknowledgment}
The authors would like to thank Prof. Michael Elad for helpful discussions and useful suggestions while writing this manuscript.
\bibliographystyle{IEEEtran}
\bibliography{references}

\end{document}

%% file: intro.tex
\section{Introduction}
Consider a noisy image
\begin{equation}
\label{eq:measurement}
 \y=\x + \n
\end{equation}
where $\y$ is the observed image, $\x$ the unknown original image
and $\n$ the contaminating noise (all in vector notation). The goal
is to reconstruct the original image $\x$ given the noisy
measurement $\y$. This is a classical formulation of image denoising, which is a typical instance of an inverse
problem. Using the maximum a posteriori (MAP) criterion, the solution  aims at maximizing the a posteriori probability given the noisy image. The MAP solution must consider prior knowledge about the distribution of $\x$, and   generally speaking, the prior distribution of natural images or any other specific class of images plays a key role in any denoising approach.

In the last few years, with the emergence of deep neural networks (DNN),  a large body of works suggests performing image denoising by feedforward neural networks when the network aims at learning image-specific or general statistics of natural images \cite{shahdoosti2019edge, tian2019deep, burger2012image, isogawa2017deep}.  Although DNN approaches are very effective and are the main focus these days, in this paper we remain loyal to the classical approaches where denoising is applied in the transform domain using shrinkage mappings. The reason for taking this position is that we are motivated by the theoretical bounds and the insights we gain by analyzing these type of approaches. 

\begin{figure}[htb]
	\centering{
	{\setlength{\fboxsep}{2pt}
		\fbox{\includegraphics[width = 0.46\textwidth]{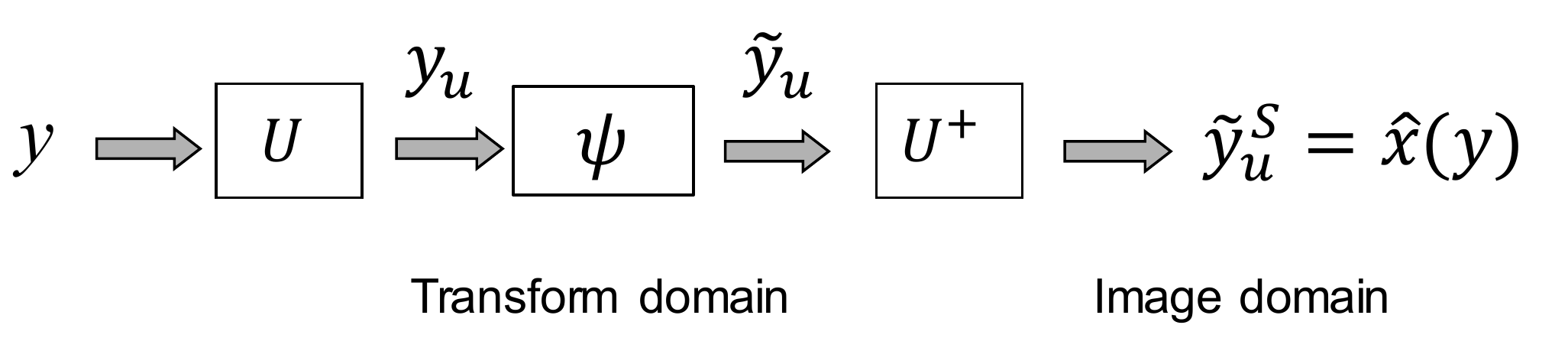}}}
		\parbox{0.7\textwidth}{
		\label{fig:shrinkage1}
 		\caption{ The shrinkage denoising in the transform domain.}
}}
\end{figure}

Transform based denoising is often
implemented using some type of wavelet transform. The main
motivation for this approach stems from the observation that the
wavelet transform of natural images tends to reduce pixel
dependencies \cite{Olshausen96a, Olshausen96b, hurri97,mallat89}.
Hence, it is possible to make a reasonable estimate about the joint
distribution of the wavelet coefficients from their marginal
distributions. When dealing with image denoising, this leads to a
family of classical techniques known as the {\it wavelet shrinkage} methods, first introduced by Donoho and Johnstone in 1994 \cite{donoho95soft, Donoho94, Donoho94b}. 
%These techniques amount to
%modifying the coefficients in the transform domain using a set of
%scalar mapping functions, $ \{ \psi_i: \Re \rightarrow \Re \}$, called {\it shrinkage
%functions} (SFs).  
The shrinkage denoising approach is composed of
a wavelet transform:
\begin{equation}
\label{eq:wt1} \y_u   = U \y
\end{equation}
where $U$ is a matrix comprising the transform basis. The transform coefficients are then rectified by a correction step in which they are modified according to a set of scalar {\it shrinkage functions}, $ \{ \psi_i: \Re \rightarrow \Re \}$:
\begin{equation}
\label{eq:sf1}
%\hat \x_{\textmd{\tiny W}} =  \psi \{
%\y_{\textmd{\tiny W}}\}
\tilde \y_u  =  \psi ( \y_u  ) \\
\end{equation}
where $\psi =(\psi_1, \psi_2,\cdots )$ is a vector of scalar
mapping functions applied to each coefficient independently:
$\tilde \y_u [i]= \psi_i( \y_u [i] )$.
The denoised image is then obtained by applying the pseudo-inverse
transform to the modified coefficients:
\begin{equation}
\label{eq:iwt1} \tilde \y_u^{\cal S} \ = U^{+} \tilde \y_u
\end{equation}
where the superscript $\cal S$ indicates that we have transformed back to the spatial (image) domain. In cases where the transform is unitary or a tight frame, the pseudo-inverse yields the adjoint; thus, $\tilde \y_u^{\cal S} \ = U^T \tilde \y_u $. The resulting image $\tilde \y_u^{\cal S}$ serves as an estimate of the original image; hence, $\hat \x(\y) = \tilde \y_u^{\cal S}$.  The denoising process is summarized in Figure~\ref{fig:shrinkage1}.

The performance of shrinkage denoising is intimately dependent on two
factors. The first factor is related to the choice of the shrinkage
functions (SF) $\{\psi_i\}$ applied to the transform coefficients. The
justification for applying a marginal (scalar) SF to each coefficient
independently emerges from the independence
assumption of the wavelet coefficients when the transform is unitary.
Assuming the statistical distribution of a wavelet band is stationary, and using the independence assumption of the wavelet coefficients, the SFs for all coefficients in
a particular wavelet band can be shown to be identical \cite{simoncelli1996noise, raphan2008optimal, simoncelli1999bayesian}.
Therefore, if the wavelet transform
$U$ is composed of $K$ bands, only $K$ SFs need to be estimated, when
$\tilde \y_u [i]= \psi_{band(i)} (\y_u [i])$, where $band(i)$ indicates the band index of pixel $i$. 
In principle, having a marginal prior distribution for a wavelet band,
the associated SF can be derived using Bayesian estimation (e.g., \cite{simoncelli1996noise, simoncelli1999bayesian}).
Alternatively, the SFs can be learnt directly from the noisy input \cite{donoho95soft, Donoho94, Donoho94b} or from a set of example images that are given
offline along with their clean counterparts \cite{hel2008discriminative, adler2010weighted, sun2015color}.

The second factor that influences the denoising performance is the
transform used during the process. Although the shrinkage approach using unitary
wavelet transforms provides good results, significant improvement is achieved
when implementing this technique with redundant transforms. Such
transforms include preselected bases such as the undecimated wavelets \cite{Coifman95},
steerable wavelets \cite{simoncelli96noise}, and other 
suggested transforms
\cite{candes99,CH2005_1006,Nezamoddini04,Do03,matalon05,Starck02}, or generated transforms
that are adaptively learnt from the noisy image \cite{elad2006image, dong2011sparsity, krim1999denoising, ouarti2009best}.
Note that scalar SFs can no longer be justified in redundant bases, as
the transform coefficients are mutually dependent due to the transform redundancy. Nevertheless, the superior
results of applying scalar SFs in the over-complete case suggest
that such a scheme is still very effective in addition to its
appealing efficiency.

The above-mentioned two factors influencing the denoising performance, namely,
the transform used and the applied SFs,
are mutually dependant and cannot be treated independently.
The type of transform used directly influences the shape of the optimal SFs.
Moreover, optimal SFs for a redundant transform, such as an undecimated wavelet,
are shown to differ from the SFs optimally designed for the unitary basis
\cite{elad2006image, raphan2008optimal, elad2006simple}. To clarify, consider finding the optimal SFs for the
unitary case with respect to the MMSE criterion. In other words, try finding a $\psi$ that minimizes
$$
\Delta = E \{ \| \hat \x(\y) -\x \|^2 \}
$$
when we have determined that $\hat \x(\y) = U^T  \psi \left( U \y \right)$, the norm $\| \cdot \|$ stands for the $\ell_2$ norm and $E \{\cdot\}$ indicates the expectation taken over $\x$
and $\y$. Whenever $U$ is unitary, this minimization can be
formulated equivalently in the transform domain (since $U U^T = U^T U = I$) as:
\begin{equation}
E \{ \| U^T  \psi \left( U \y \right) -\x \|^2 \} = E \{ \| \psi \left(
U \y \right) -U \x \|^2 \}
\end{equation}
i.e., $\psi$  is optimized so that the noisy transform coefficients $U\y$  should be as close as possible to the transform coefficients of the clean image $U\x$. For an over-complete transform, however, this equality is no longer valid (since $U U^T \neq I$). This implies that
the optimization for $ \psi$ should be expressed in the spatial
domain, which is the relevant domain in our case.  Because the inverse transform couples wavelet
coefficients (inside subband and between subbands), spatial domain optimization requires a joint minimization of all SFs simultaneously, and this optimization is far more complicated to apply.
In fact, this might be the reason that SFs applyed in redundant
bases are commonly borrowed from the unitary case or optimized in
the transform domain with no real justification.

In \cite{raphan2008optimal}, Raphan and Simoncelli  showed that as long as the statistics of the image and the noise are stationary, the expected MSE of the denoised image resulting from applying the SFs, $\psi$ in the unitary bases, is always greater or equal to the MSE of the denoised image resulting
from applying the same $\psi$ in the redundant basis
(by spatially replicating the unitary basis using, e.g., cycle spinning or undecimated subbands \cite{Coifman95, starck2007undecimated} ). Note that this property was proven irrespective of the type of applied SFs, $\psi$. They also showed that when working with a redundant basis, there is an advantage in optimizing the SFs (with respect to the expected error) in the spatial domain rather than in the transform domain. This requires, however, optimizing jointly all SFs simultaneously, making the optimization process a demanding task.

In this paper, we extend the results of  \cite{raphan2008optimal} and establish a complete picture of the interrelations between the transform used, the optimal shrinkage functions, and the domains in which they are optimized. In particular, we show that  for subband optimization, where each $\psi_i$ is optimized independently, optimizing each SF in the spatial domain is always better than or equal to optimizing the SFs in the transform domain. This option, besides being simple to implement, is proven to outperform
the traditional transform domain optimization while avoiding the demanding spatial domain optimization of all SFs simultaneously. 

Additionally,  for redundant bases, we provide the expected denoising gain we may achieve, relative to the unitary basis, as a function of the basis redundancy. This result allows a user to make a clever decision about the redundancy used by taking into account the expected denoising gain and the computational time allocated for this process.

%\begin{figure}[htb]
%	\centering{
%		\includegraphics[width = 0.6\textwidth]{Figures/relations.pdf}
%		\parbox{0.8\textwidth}{
%		\label{fig:relations}
% 		\caption{ The expected MSE denoising performance when optimizing the SFs in unitary v.s. redundant bases, and spatial v.s. transform domains.}
%}}
%\end{figure}
%

%% file: redundant.tex
\section{Redundant vs. Unitary Transforms}
\label{sec:redundant}
A common axiom in image denoising is that denoising applied in
redundant bases (cycle-spinning or undecimated wavelets) outperforms the results
obtained in unitary transforms \cite{Coifman95, raphan2008optimal}. 
In this section we examine the relationships between the unitary and redundant transforms. In particular, we offer theoretical justification
for applying shrinkage denoising in redundant bases. Some of the relationships in this section were already proven in \cite{raphan2008optimal} but we repeat them here for clarity and to provide a complete picture of the interrelations between the transformed used and the domain where the MSE is optimized. 
To be able to compare different transforms on a common basis, we limit our discussion to the unitary basis and their corresponding cycle-spinning transforms. That is to say, the difference between a unitary and a redundant transform is that the former is properly decimated and thus forms a complete basis, while the latter is formed by cycle-spinning the unitary basis.

A common technique for shrinkage denoising in a redundant basis is the cycle-spinning framework \cite{Coifman95}. Cycle-spinning is performed by applying a unitary transform on a set of shifted versions of the image, denoising each version independently, then
averaging the results after properly shifting them  back. Since the transform of a spatially shifted image can be applied equivalently by shifting the transform basis
(by the same amount but in the opposite direction), the transform can be seen as
a redundant transform, composed of a set of shifted versions of the original unitary transform:
\begin{equation}
\label{eq:wi}
\y_{u_i}  = U S_i  \y = U_i \y  ~~~i=1 \ldots N
\end{equation}
where $S_i$ is a (cyclic) shift operator by the $i$-th displacement, and $U_i=U S_i$ is a unitary transform composed of the wavelet basis after applying
the respective shift. The entire transform is constructed by concatenating together
all shifted transforms:

\begin{equation}
\label{eqn:undecimatedW} 
\y_{w}  =  W \y
\end{equation}
where the redundant transform is defined as follows:
\begin{equation}
\label{eqn:DefT1} {W} = \frac{1}{\sqrt{N}} \left [
\begin{array}{c}
U_1 \\
U_2 \\
\vdots \\
U_N
\end{array}
\right ]
\end{equation}
Note the $W$ is over-complete and tight frame, satisfying $W^T W = I$; however, $W W^T \neq I$. Since
$W W^T$ is a projection matrix\footnote{A square matrix $A$ is a projection matrix iff $AA=A$.}, it can be shown that
this restricts the eigen-values of $W W^T$ to be $1$ or $0$.
If $W$ is an $m \times n$ matrix ($m>n$), then there are $n$ eigen-values of $1$, and
$m-n$ eigen-values of $0$ \cite{golub2012matrix}.
Consequently, for any $n \times 1$ vector  $\x$ and $m \times 1$ vector $\z$, we have:
\begin{equation}
\label{eq:tight1}
\| W \x \| = \| \x \|
\end{equation}
and
\begin{equation}
\label{eq:tight2}
\| W^T \z \| \leq \| \z \|
\end{equation}
Similarly, since $U_i$ is unitary, we have $U_i^T U_i = U_i U_i^T =I$ and accordingly:
\begin{equation}
\label{eq:uni}
\| U_i \x \| = \| U_i^T \x \| = \| \x \|,  ~~~\forall ~i
\end{equation}

Denote by $\n=\x-\y$ the contaminated noise in the image domain before denoising.
By applying a unitary transform, the error is transformed as well:
$$
U (\x - \y) = U \n \doteq \n_u 
$$
and similarly, by the redundant transform:
$$
W (\x - \y) = W \n \doteq \n_w 
$$
Since the transforms $U$ and $W$ are tight frames, we have (following Equations~\ref{eq:tight1} and~\ref{eq:uni}):
\begin{equation}
\label{eq:n_w}
\| \n \| = \| \n_u  \| = \| \n_w  \|
\end{equation}
which means that the norm of the error in the transform domain
equals its norm in the image domain, and this is true for the unitary as well
as the redundant case.

Now, after applying the shrinkage functions $\psi$ to the transform coefficients, the distortion value may change. We define:
$$
\tilde \n_u  = U \x - \psi(U \y)  ~~\mbox{and similarly}~~
\tilde \n_w = W \x - \psi(W \y)
$$
For the unitary case, the distortion is propagated to the image domain via the inverse transform:
$$
 U^T (U x - \psi(U \y)) = U^T  \tilde \n_u  \doteq \tilde \n_u^{\cal S}
$$
and following Equation~\ref{eq:uni}, we have:
\begin{equation}
\label{eq:wtd}
\| \tilde \n_u   \| = \| \tilde \n_u^{\cal S}  \|
\end{equation}
i.e., after applying the SFs, the MSE distortion in the transform domain is identical to its distortion in the image domain. As we will see next, in redundant transforms this property is not satisfied. In redundant transforms, the error in the image domain is:
$$
W^T (W x - \psi(W \y)) = W^T \tilde \n_w  \doteq  \tilde \n_w^{\cal S}
$$
Nevertheless, following Equation~\ref{eq:tight2}, we have:
\begin{equation}
\label{eq:utd}
\| \tilde \n_w  \|  \geq \| \tilde \n_w^{\cal S} \|
\end{equation}
Note that the above relations (Equations~\ref{eq:utd} and \ref{eq:wtd}) are valid for {\em any} shrinkage functions $\psi$ and for any $\x$ and $\y$.

\begin{figure}[bth]
\label{fig:relations} 
\centering{
\includegraphics[width=0.35\textwidth]{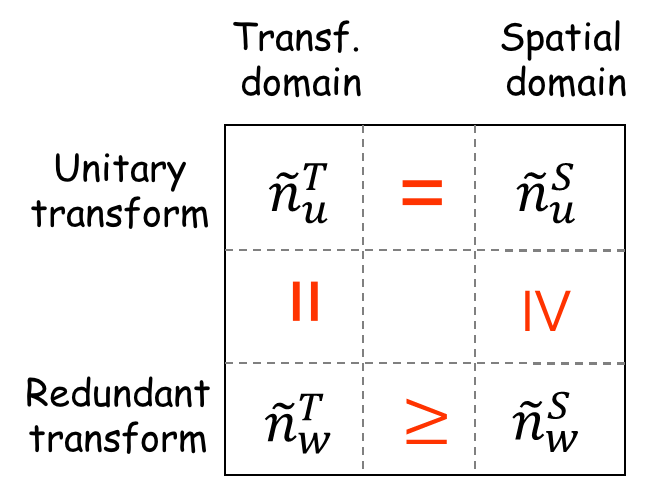}
 \centering {\parbox{0.47\textwidth}{
 \caption{ The expected remaining noise for unitary vs. redundant bases, and spatial vs. transform domains.}
 }}}
\end{figure}

Letting $\s$ be a vector value depending on $\x$ and $\y$, we define the expected RMSE of $\s$:
$$
\left \| \s \right \|_E  \doteq  \sqrt { E \left  \| \s \right \|^2 }  
~~~~\mbox{where}~~~~ 
E \left  \| \s \right \|^2  = \int \left  \| \s  \right  \|^2 P(\x,\y) d\x d\y
$$
Since relations \ref{eq:utd} and \ref{eq:wtd} are true for any $\x$ and $\y$, we can rephrase these relations using a statistical point of view:
\begin{equation}
\label{eq:wtd1}
\| \tilde \n_u   \|_E = \| \tilde \n_u^{\cal S}  \|_E
\end{equation}
and
\begin{equation}
\label{eq:utd1}
\| \tilde \n_w  \|_E  \geq \| \tilde \n_w^{\cal S} \|_E
\end{equation}
These relations are illustrated along the two rows of Figure~\ref{fig:relations}.
We now establish the relationships between the unitary and redundant transforms that are indicated in the two columns of Figure~\ref{fig:relations}.

We first show that in the transform domain, for the two transforms, the expected MSE distortion is
equal. This outcome stems from the stationary property of natural images, where it is assumed that the statistical properties of natural images are shift invariant.
\begin{theorem}
After denoising, the expected MSE distortions in the transform domain are equal for the unitary and for the redundant transforms. In other words, for any given $\psi$:
$$
\left \| \tilde \n_u   \right \|_E = \left \| \tilde \n_w   \right \|_E
$$
\end{theorem}

\noindent
{\bf Proof 1:} In Appendix A.

The last theorem leads to a theoretical justification for applying shrinkage denoising in a redundant basis. This is explicitly expressed in the next theorem:

\begin{theorem}
For any given $\psi$,
$$
 \| \tilde \n_u^{\cal S}  \|_E  \geq
 \| \tilde \n_w^{\cal S}  \|_E
$$
\end{theorem}
\noindent
{\bf Proof 2:} In Appendix B.

Theorem~2 completes the entire picture of Figure~\ref{fig:relations}:  In the transform domain, the expected remaining noise, after shrinkage,  is identical for the unitary and the redundant (cycle-spinning) wavelet transforms for {\em any} shrinkage functions. 
When transforming back into the spatial domain, however, the remaining noise is expected to decrease in the redundant transform while staying the same in the unitary transform. This main conclusion suggests that it is preferable to apply shrinkage denoising in a redundant basis rather than in the unitary basis. 
%In Section \cite{XX} we provide a quantitative justification for this conclusion where we provide the expected MSE gain for image denoising as a function of the redundancy used.

%% file: bounds.tex
\section{Optimizing the Shrinkage Functions}
In this section we deal with the objectives to which the shrinkage functions (SFs) are optimized. As mentioned earlier, SFs play a significant role in the resulting performance, and their optimization is a longstanding topic of study (see, e.g., \cite{hyvarinen1999sparse, elad2006image, moulin1998analysis, xiao2011comparative}, just to name a few).  In principle, SFs can be derived from the
joint statistics of the transform coefficients \cite{moulin1998analysis, hyvarinen1999sparse, raphan2008optimal}, but, unfortunately, modeling the precise joint statistics is a complicated and still intractable problem.  Alternatively, one can optimize the SF of each subband independently using marginal statistics, but as mentioned earlier, this is not optimal in the redundant case. Another option is to learn the optimal  SFs from an ensemble of  images using a set of noisy and clean examples  \cite{sliceTR06, raphan2008optimal}, where the SFs are designed to clean the noisy examples in an optimal manner towards their clean counterparts. As shown next, there are several domains in which the SFs can be optimized in and the resulting quality depends on the selected domain.

\begin{figure}[bth]
\centering{\
{\setlength{\fboxsep}{10pt}\fbox{\includegraphics[width=0.46\textwidth]{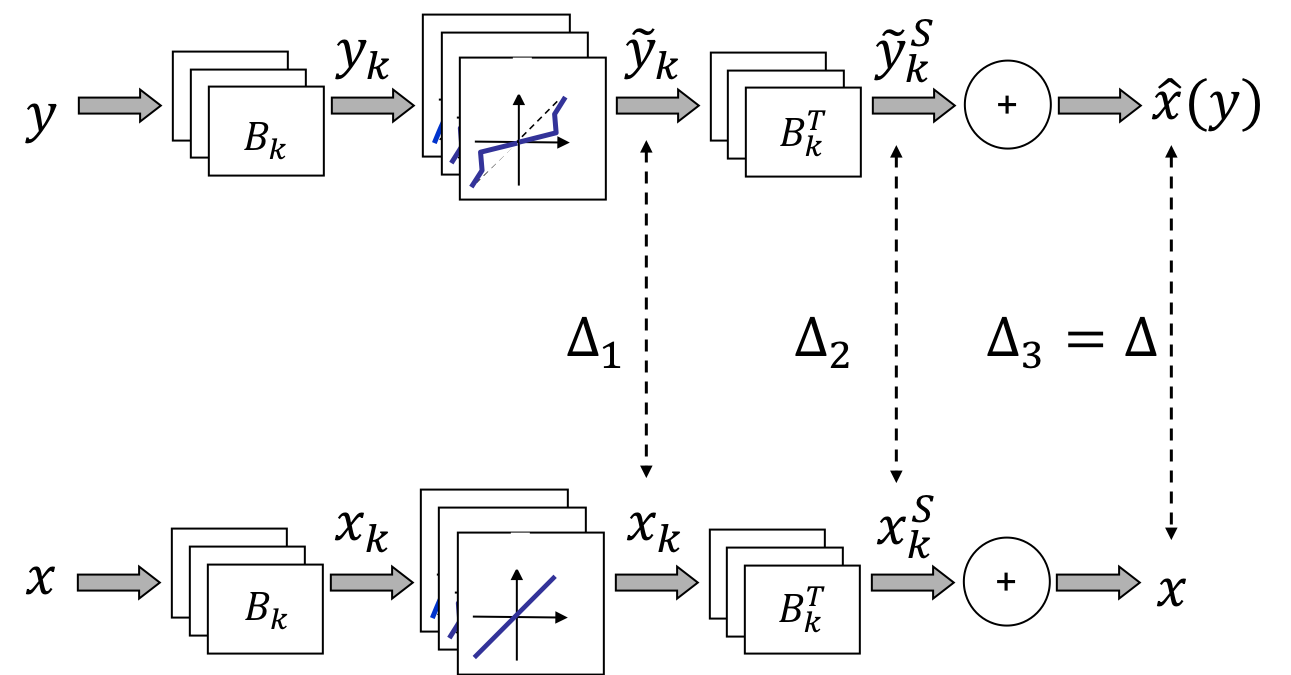}}}
 \centering {\parbox{0.5\textwidth}{
 \label{fig:shrinkComp} 
 \caption{The  denoising process and three optimization schemes for the shrinkage functions.}
 }}}
\end{figure}

In undecimated wavelet transforms, the number of
coefficients is $K$ times the size of the image, where $K$ is the
number of the wavelet bands. To facilitate the notation for band operations, we reorder the rows of
a transform $W$ so that transform rows corresponding to a wavelet band are co-located in a
block. Naturally, we extend the same reordering to $\y_w$.
Assuming we have $K$ different wavelet bands and a corresponding
permutation matrix $P$:
\begin{equation*}
\label{eqn:DefB}
B =P W = \left [
\begin{array}{c}
B_1 \\ B_2 \\ \vdots \\ B_K
\end{array}
\right ] ~~\mbox{and accordingly}~  \y_{B}  = B \y = \left [
\begin{array}{c}
\y _{1} \\ \y _{2} \\
\vdots \\ \y _{K}
\end{array}
\right ]
\end{equation*}
where $\y _{k}=B_k \y$ represents the coefficients of the $k^{th}$ band.
The new reordering does not change the tight frame property;
thus, if $W^T W = I$, we have $B^T B = I$ as well.
In the new reordering, a vector of SFs, ${\psi
}=[{\psi}_1, \psi_2, \cdots, \psi_K]$, can be represented
efficiently as follows in Equation~\ref{eq:phi22}.
Since $\psi_k$ is applied similarly to
all coefficients in the $k^{th}$ band,
we can rewrite Equation~\ref{eq:sf1} as
\begin{equation}
\label{eq:phi22}
\tilde \y _{k} = \psi_k ( \y _{k} )
\end{equation}
which means that the scalar mapping $\psi_k: \Re \rightarrow \Re$
is applied individually to each entry in  $\y _{k}$.
The clean image is then estimated using the adjoint:
\begin{equation}
\label{eqn:OverCompB}
\hat \x(\y) = B^T \psi (\y _B) = \sum_{k=1}^K
B_k^T  \tilde \y _{k} \doteq \sum_{k=1}^K
\tilde \y^{\cal S}_{k}
\end{equation}
where we define $\tilde \y^{\cal S}_{k} = B_k^T  \tilde \y _{k}$.
This process is illustrated in the upper pipeline of Figure
\ref{fig:shrinkComp}.
 Let the SFs be a set of mapping functions taken from a given function
space $\Psi$.
The optimal set of SFs with respect to the MSE criterion
is then obtained by finding the function set
$\psi \subset \Psi$
that minimizes the following objective:
%Let the SFs to be approximated using parametric functions
%$\psi_k(s) \approx \psi_k(s ; \p_k) \}$, where $\p_k$ is a parameter vector
%defining $\psi_k$. The optimal set of SFs with respect to MSE criterion
%is then obtained by finding a global parameter vector
%$\p = \left [ \p_1,\cdots,\p_K \right ] $
%that minimizes the following objective:
$$
\hat \psi= \arg \min_{\psi \in \Psi}  \Delta(\psi)
$$
where
\begin{equation}
\Delta ({{ \psi}}) = \sqrt{E \left \{ \| \x- \hat \x(\y ) \|^2 \right \}}
\label{eq:delta}
\end{equation}
where $\hat \x(\y )$ is as estimate defined in Equation~\ref{eqn:OverCompB}
and $E \{\cdot\}$
stands for the expectation taken over $(\x ,\y)$.

The above minimization is complicated to accomplish as it requires
modeling the entire joint statistics of natural images.
 Below we consider other alternatives for the objective functions.
We examine three objectives expressing the optimal set of
SFs. We refer to the definitions illustrated in Figure~\ref{fig:shrinkComp}.

\begin{itemize}
\item
{\bf Method 1} (transform domain -- independent bands):
A set of SFs are optimized in the transform domain. The
optimization is applied by minimizing the
objective function:
$$
\Delta_1= \sum_k \sqrt{E \left\{ \left \| \x _{k} - \tilde \y _{k} \right \|^2 \right \}}
$$
where $\x _{k}= B_k \x$ is the clean counterpart of $\tilde \y _{k}$. Since this objective is composed of
a sum of independent terms, each of which contains a particular band,
the minimization of this objective can be applied at each
wavelet band independently, using only its marginal statistics (since the SFs are scalars); namely:
$$
\hat \psi_k= \arg \min_{\psi \in \Psi}  E \left\{ \left \| \x_{k} - \tilde \x_{k} \right \|^2 \right \}~~~\forall~k \in \{ 1,2,..,K\}
$$
where the expectation is over $\y_{k}, \x_{k}$.

%Note, that the marginal statistics is sufficient also if we
%do not assume the independence assumption of the wavelet coefficients.
%In fact, as long as the denoising algorithm is composed of
%scalar SFs that are applied to each coefficient independently, only marginal
%statistics are required for the minimization.
\item
{\bf Method 2} (spatial domain -- independent bands):
A set of SFs is optimized in the spatial domain. The objective term
for this method reads:
$$
\Delta_2=\sum_k  \sqrt{E \left\{ \left \| \x^{\cal S}_{k} - \tilde \y^{\cal S}_{k} \right \|^2 \right \}}
$$
where $\x^{\cal S}_{k}=B^T_k \x _{k}$ and $\tilde \y^{\cal S}_{k} = B_k^T \tilde \y _{k}$ .
Note that though the objective criterion is expressed in the spatial
domain, the SFs can be optimized for each band independently.
Thus, although intra-band dependencies are conveyed through the
adjoint transform and must be considered, the inter-band
dependencies are ignored.
\item
{\bf Method 3} (spatial domain -- joint bands):
The objective goal is expressed in the spatial domain:
$$
\Delta_3=\sqrt{ E \left\{\left \| \sum_k ( \x^{\cal S}_k - \tilde \y^{\cal S}_k) \right \|^2 \right \}}
$$
It is easy to verify that this objective gives the actual expected error
as defined in Equation~\ref{eq:delta}; thus, $\Delta_3=\Delta$.
In this scheme, the SFs are evaluated simultaneously
while inter-band as well as intra-band dependencies must be taken into account.
\end{itemize}

\noindent Denote the deviation of the approximated coefficients from the clean coefficients by  $\d_k=\x_k  - \tilde \y_k $ and similarly
$\d_{k}^{S}=B_k^T(\x_k  - \tilde \y_k )= B_k^T \d_k$.
Using this notation, the above objectives read:
\begin{equation}
\Delta_1 = \sum_k  \sqrt { E \left \{ \left \| \d_{k} \right \|^2 \right \}}~~, ~~\Delta_2 = \sum_k  \sqrt { E \left \{   \left \| \d_{k}^{\cal S} \right \|^2 \right \}}~, 
\end{equation}
and
$$
\Delta_3  = \sqrt { E \left \{  \left \| \sum_k \d_{k}^{\cal S} \right \|^2 \right \} }= \Delta 
$$

%For the sake of clarity we also define
%a squared versions of the objective functions:
%\begin{equation}
%\Delta'_1 = E \left \{ \sum_k \left \| \d_{c_k} \right \|^2 \right \} ~~, ~~~\Delta'_2 = E \left \{ \sum_k  \left \| \d_{k} \right \|^2 \right \} ~~, ~~~ \Delta'_3 =  E \left \{ \left \| \sum_k \d_{k} \right \|^2 \right \}
%\end{equation}
%The squared objectives are not simply the squared values of the original objectives, but they share the optimal shrinkage functions.
%The following theorem shows that the optimal shrinkage function obtained for $\Delta_i$ is similar to the optimal shrinkage function obtained for $\Delta'_i$, $~i = 1..3$:

\begin{figure}[tbh]
\centering{
\includegraphics[width=0.45\textwidth]{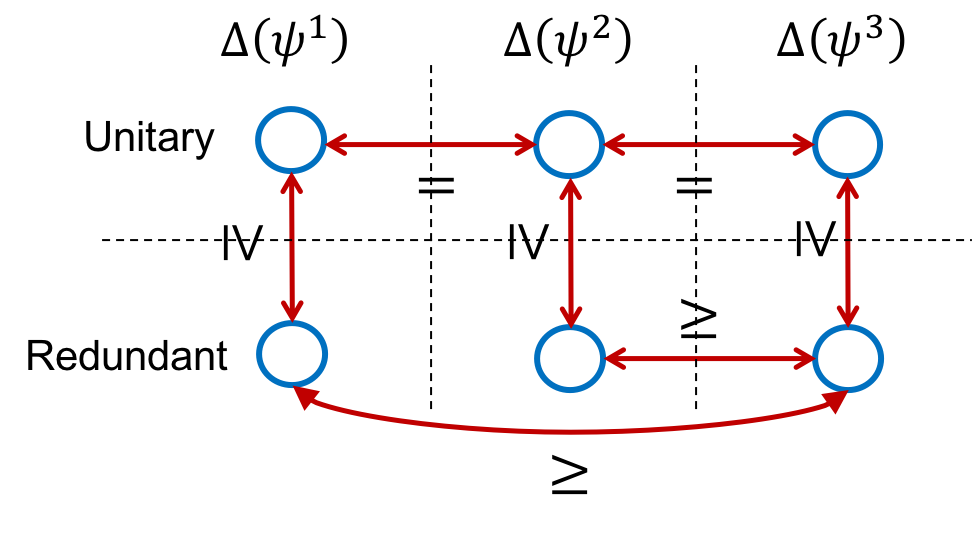}
 \centering {\parbox{0.5\textwidth}{
 \caption{
 \label{fig:table}
 The relationships of the expected error, $\Delta(\psi)$,  when
 applying shrinkage functions that have been optimized using penalty 1..3 and
 in unitary vs. redundant transforms.}
  }}}
\end{figure}

For each method defined above, denote  an associated
optimal SF, ${\hat \psi}^i$, as follows:
\begin{equation}
\label{eq:opt_psi}
{{\hat \psi}}^i = \arg \min_{ \psi \in \Psi} \Delta_i , ~~~\mbox{for}~~i=1..3
\end{equation}
Additionally, the objective $\Delta(\hat \psi^i)$ denotes the actual expected error as defined in Equation
~\ref{eq:delta}  when applying the SF $\hat \psi^i$.
In the following, we show that if the wavelet transform is unitary, then all three methods produce the same result. This is illustrated in the upper line of Figure~\ref{fig:table}.

\begin{theorem}
For the unitary case we have:
$$
\Delta(\hat \psi^1) = \Delta(\hat \psi^2) = \Delta(\hat \psi^3)
$$
\end{theorem}

\noindent
{\bf Proof 3}:
To show the above relations, we prove that actually, for the unitary case,
$\hat \psi^1=\hat \psi^2=\hat \psi^3$,  which derives the theorem.
Recall that $\hat \psi=[\hat \psi_1 \cdots \hat \psi_K]$ is composed of $K$ SFs, each of which applies to a particular band. Thus, $\d_k$ and $\d_{k}^{\cal S}$ depend only on $\psi_k$, and we can apply the optimization to each band independently. For the $k^{th}$ band, we have:
$$
{\hat \psi}^1_k = \arg \min_{\psi_k} E \left \{ \left \| \d_{k} \right \|^2 \right \} $$
and
$$
{\psi}^2_k = \arg \min_{\psi_k} E \left \{ \left \| B_k^T \d_{k} \right \|^2 \right \} 
$$
Since $W$ is unitary, $WW^T=I$, and accordingly, $B_i B_j^T=\delta_{i,j} I$.
Using this relation we get:
$$
\left \| B_k^T \d_{k} \right \|^2 = \left \| \d_{k} \right \|^2
$$
which gives
$$
\hat \psi^1_k =  \hat \psi^2_k,
~~~~\mbox{for}~~k=1..K
$$
and accordingly $\hat \psi^1=\hat \psi^2$, which implies the first relation in the theorem.

Similarly,
$$
\hat \psi^3_k=\arg \min_{\psi_k} E \left \{ \left \| \sum_j  B_j^T \d_{j} \right \|^2 \right \}
$$
However,
$$
\left \| \sum_j  B_j^T \d_{j} \right \|^2 =  \left ( \sum_i  \d_{i}^T B_i \right )
\left ( \sum_j  B_j^T \d_{j} \right )  =  \sum_j  \left \| \d_{j} \right \|^2
$$
Thus,
\begin{eqnarray} \nonumber
\hat \psi^3_k=\arg \min_{\psi_k} E \left \{ {  \sum_j  \left \| \d_{j} \right \|^2 }  \right \} = \arg \min_{\psi_k} E \left \{ \left \| \d_{k} \right \|^2   \right \}= \hat \psi^1_k
\end{eqnarray}
where $k \in \{1..K\}$. This yields $\hat \psi^3=\hat \psi^1$, which implies the second relation of the theorem.
Hence, in the unitary case, optimizing the SFs using any one of the above methods is equivalent~~~$\Box$.

%Defining a concatenation of all
%$\{ \d_{c_k} \}$ into a single vector as $\d_c$ and a concatenation of all
%$\{ \d_{k} \}$ as $\d$:
%$$
%\d_c=
%\left (
%\begin{array}{c}
%\d_{c_1} \\ \d_{c_2} \\ \vdots \\ \d_{c_K}
%\end{array}
%\right )~~~\mbox{and}~~~
%\d =
%\left (
%\begin{array}{c}
%\d_{1} \\ \d_{2} \\ \vdots \\ \d_{K}
%\end{array}
%\right )
%$$
%Note, that $C^T \d_c = \d$. Since $C C^T=I$ we have

\begin{figure}[bth]
\centering{\
\includegraphics[width=0.5\textwidth] {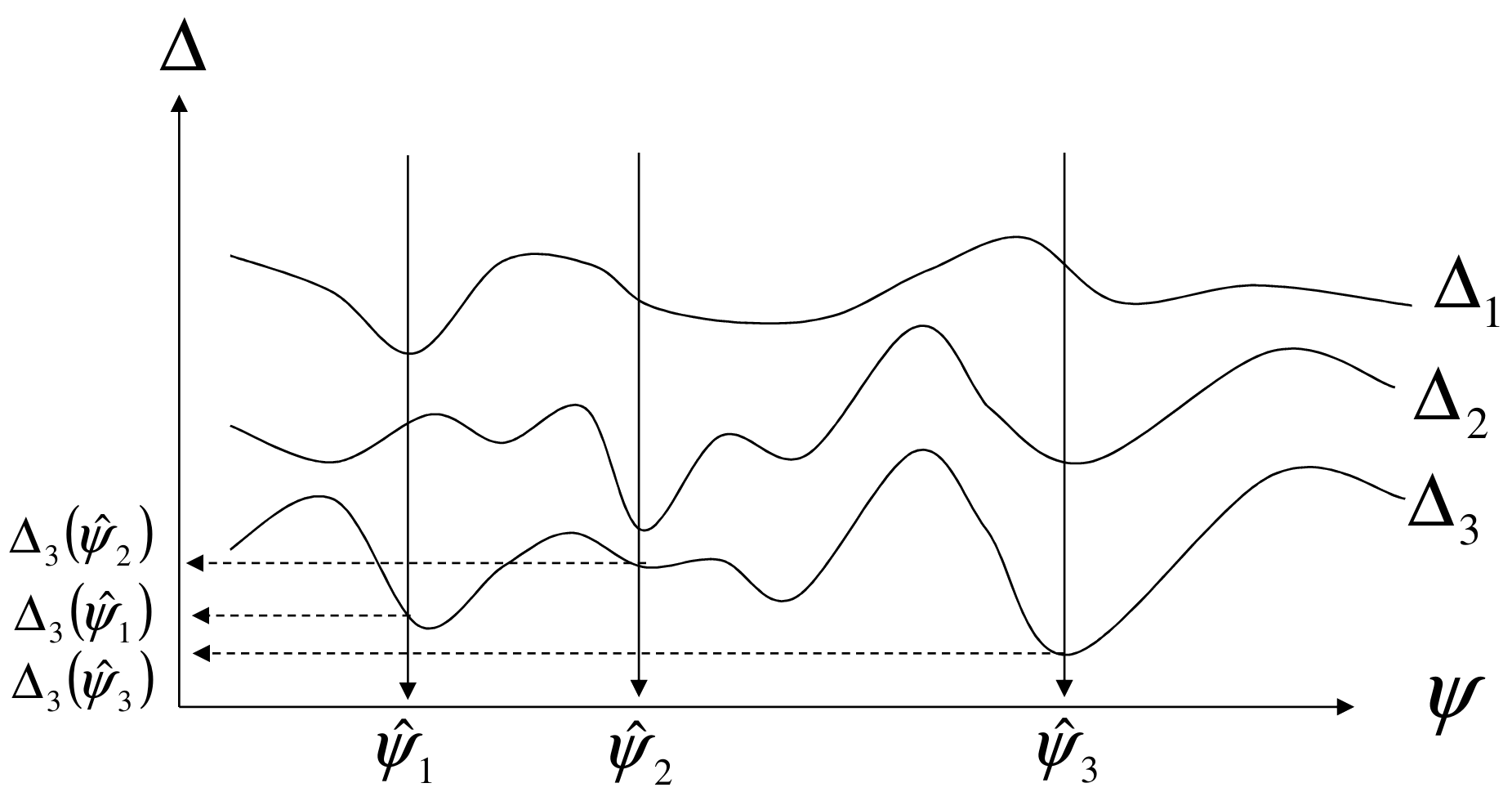}}
\centering {\parbox{0.45\textwidth}{
 \caption{
 \label{fig:deltas} Illustrated profiles of penalties following Theorem \ref{th:4}. The x-axis indicates various $\psi$ values. Note that in this example, $\Delta(\hat \psi^1)  <  \Delta(\hat \psi^2) $   }}}
\end{figure}

\vspace{10pt}
Theorem~3 establishes the justification for optimizing the
SFs in the transform domain, in cases where the transform used is unitary.
Using Method~1, each individual SF can be optimized independently,
collecting only marginal statistics. This property makes this scheme
very appealing and thus very popular (e.g., \cite{donoho95soft,Coifman95,simoncelli96noise}).
Theorem~4 shows that in the over-complete transform, the
situation is totally different, and the domain in which we apply the optimization
makes a difference (see Figure \ref{fig:deltas} for an illustration).

\begin{theorem}
\label{th:4}
Let the transform $W$ be over-complete and tight frame.
In such a case, for each~$\psi$,
$$
\Delta_1(\psi) \geq \Delta_2 (\psi) \geq \Delta_3(\psi)
$$
\end{theorem}

\noindent
{\bf Proof 4}:  
Since $W$ is tight frame, it follows that $W^T W =I$. It can easily be shown that
this restricts the norm of each $B_k$: $\| B_k \| = \lambda_{(k)} \leq 1$, where
$\lambda_{(k)}$ denotes the maximal eigen-value of $B_k B_k^T$ \cite{golub2012matrix}.
This yields that for any vector $\z$, $\| B_k^T \z \| \leq \| \z \|$; hence,
\begin{align*}
\Delta_1 &= \sum_j \sqrt{E \left \{  \| \d_{j} \|^2 \right \}}
\geq \sum_j \sqrt{ E \left \{  \| B_j^T \d_{j} \|^2 \right \}} \\
&= \sum_j \sqrt{ E \left \{ \| \d_{j}^{\cal S} \|^2 \right \}} = \Delta_2 
\end{align*}
and this proves the first inequality.
Due to the triangular inequality of a norm\footnote{Note that the expectation value can be inserted into the norm definition.},
we also have:
$$
\Delta_2 = \sum_j \sqrt{ E \left \{ \| \d_{j}^{\cal S} \|^2 \right \} } \geq \sqrt{ E \left \{ \left \| \sum_j  \d_j^{\cal S} \right \|^2 \right \} } = \Delta_3
$$

%Denote $\d = \sum_k \d_k$ and $\d_c$ as a vector concatenating
%together all $\d_{c_k}$ into a single vector:
%$$
%\d_c=
%\left (
%\begin{array}{c}
%\d_{c_1} \\ \d_{c_2} \\ \vdots \\ \d_{c_K}
%\end{array}
%\right )
%$$
%We have that $\d = H \d_c$ where $H$ is a matrix concatenating
%$K$ identity matrices: $H=\left [ I, I, \cdots, I \right ]$.
%Since we have $\| H^T H \| = {K}$ it follows that for any vector $\z$,
%$\| H \z \|^2 < K \| \z \|^2$ which gives
%$$
%\Delta_3 = \| H \d_c \|^2 \leq K \| \d_c \|^2  = K \sum_k \| \d_{c_k} \|^2 = K \Delta_2
%$$

which gives the second inequality in the theorem~~~$\Box$.

%\section{Optimization in Practice}

%As explained in the last section, finding the optimal shrinkage functions can be performed either in the transform domain (method 1), requiring collecting only marginal statistics, or in the spatial domain (method 3), where joint statistics is required.
%Theorem~1 states that, in the unitary case, either way produce the same results.
%This is probably the reason why the traditional approaches optimize the SFs
%in the transform domain (e.g. \cite{donoho95soft,Coifman95,simoncelli96noise}).

Theorem~\ref{th:4} reveals that in the over-complete case, optimizing the SFs in the transform domain
is not optimal. In the following we provide justification for optimizing the SFs using Methods 1 and 2 as they provide upper bounds for the desired penalty ($\Delta$ of Method 3), that might be difficult to achieve.

\begin{theorem}
\label{th:5}
Let $\hat \psi^i = \arg \min_{\psi \in \Psi} \Delta_i(\psi)$ as defined in Equation~\ref{eq:opt_psi}. In the over-complete case,
\begin{equation}
\label{eq:ineq2}
\Delta_1(\hat \psi^1) \geq  \Delta_2(\hat \psi^2)  \geq  \Delta_3(\hat \psi^3)
\end{equation}
\end{theorem}
\noindent
{\bf Proof 5:}
The SF ${\hat{ \psi}}^3$ minimizes $\Delta_3$; thus,  $\Delta_3 (\hat \psi^2)  \geq \Delta_3 (\hat \psi^3)$.
Following Theorem~\ref{th:4}, however, we have that $\Delta_2 (\hat \psi^2) \geq \Delta_3 (\hat \psi^2) $, from which it readily follows  that $\Delta_2 (\hat \psi^2) \geq \Delta_3 (\hat \psi^3)$.
The second inequality can be shown using a similar argument. Q.E.D. \\

Note also that according to
the proof above, the actual errors (i.e., $\Delta_3=\Delta$) using $\hat \psi^2$ and $\hat \psi^1$ are even
tighter, i.e.:
\begin{eqnarray}
\label{eq:c32}
\Delta_2(\hat \psi^2)  \geq \Delta(\hat \psi^2) \geq \Delta(\hat \psi^3) ~~~\mbox{for Method 2}\\
\label{eq:c31}
\Delta_1(\hat \psi^1)  \geq \Delta(\hat \psi^1) \geq \Delta(\hat \psi^3) ~~~\mbox{for
Method 1}
\end{eqnarray}
and since $\Delta_1(\hat \psi^1)  \geq  \Delta_2(\hat \psi^2)$ (Theorem~\ref{th:5}), the
SF $\hat \psi^2$ has a better bound
than $\hat \psi^1$.  Thus, it is {\em expected} that $\Delta(\hat \psi^1)  \geq  \Delta(\hat \psi^2)$.
Nevertheless, it cannot be assured that the actual error for $\hat \psi_2$
outperforms the actual error of $\hat \psi_1$, i.e., the relation:
\begin{equation}
\label{eq:c12}
\Delta(\hat \psi^1)  \geq  \Delta(\hat \psi^2)
\end{equation}
is not necessarily true. To prove this, see a counter-example in Figure~\ref{fig:deltas}.

To conclude, in redundant transforms $\Delta_3=\Delta$ determines the actual error and it is the optimal penalty to minimize. Nevertheless, since it requires inter- and intra-bands statistics, it is sometimes complicated to optimize. $\Delta_1$ is the easiest term to minimize as it requires collecting only marginal statistics. Indeed, this approach is commonly used in the traditional techniques (hard/soft thresholding originated from this penalty).
$\Delta_2$ is a better penalty to minimize than $\Delta_1$ as its bound is tighter, although it might be harder to optimize as it requires modeling intra-band statistics. Nevertheless,  it is not guaranteed that $\Delta(\hat \psi^1) \geq  \Delta(\hat \psi^2)$. Thus, there is an inherent trade-off between the three methods, while spatial domain optimization (Method 3) is preferable with respect to denoising quality, transform domain optimization (Method 1) is the most efficient to apply. Weak spatial domain (Method 2) is a good compromise between quality and efficiency.
% (see Figure~\ref{fig:quality} for an illustrative summary).

In unitary transforms, all optimization objectives (Methods 1, 2, and 3) will generate similar results. Using Theorem~2 above, however, it was proven that it is expected that denoising in redundant transforms will generate better results than using unitary transforms. These relations are summarized in Figure~\ref{fig:table}.

\section{Improvement Rates for Redundant Transforms} 
\begin{figure}[htb]
\centering{
\includegraphics[width=0.5\textwidth]{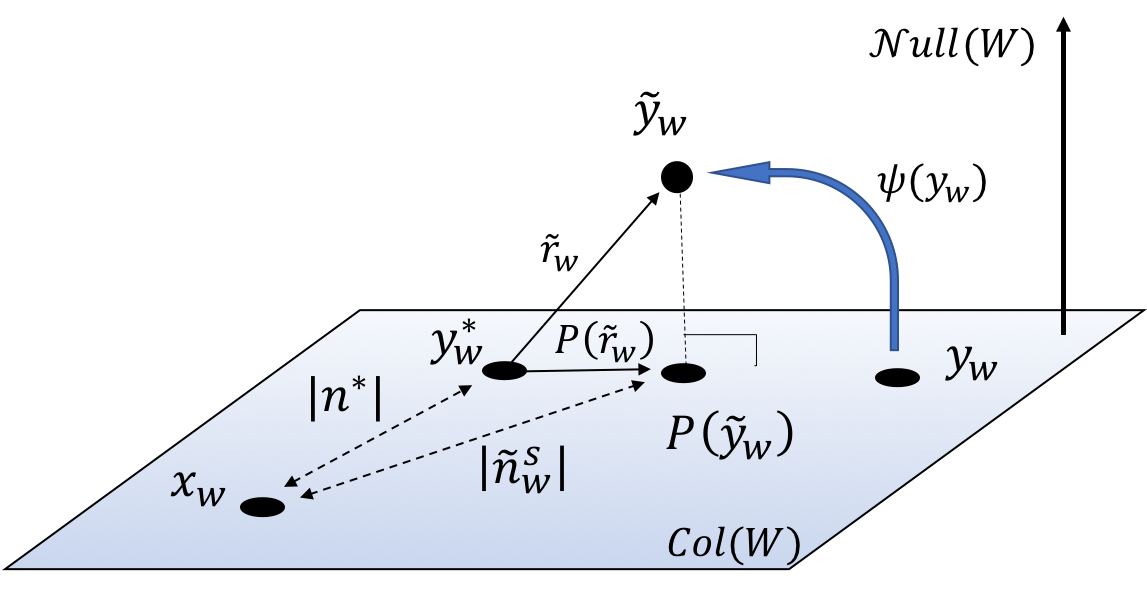}
 \centering {\parbox{0.5\textwidth}{
 \caption{
 \label{fig:proj_y}
 Due to the triangular inequality, $\|\tilde \n_w^S \| \leq \| \n^*\| + \| P(\tilde \r_w)\|$. Since $\| P(\tilde \r_w) \|_E =  \frac{1}{\sqrt{k}} \|\tilde \r_w\|_E$, Theorem~6 follows. }
  }}}
\end{figure}

In this section we analyze the expected improvement of the remaining MSE with respect to the redundant rate of the transform used. We assume an over-complete wavelet transform made by cycle spinning as given in Equation \ref{eqn:undecimatedW}:
\begin{equation}
\y_{w}  =  W \y
\end{equation}
where $W$ is composed of $k$ shifted versions of the unitary transform $U$.
\begin{equation}
\label{eqn:DefT2} 
{W} = \frac{1}{\sqrt{k}} \left [
\begin{array}{c}
U_1 \\
U_2 \\
\vdots \\
U_k
\end{array}
\right ]
\end{equation} 
The redundancy rate of this transform is $k$, where $k\in \{ 1 \cdots n\}$. Namely, each subband has $k$ shifts of the corresponding basis function.  Note that if $k=1$, the transform reduces to the unitary transform while the maximal redundancy is when $k=n$. The transform is composed of a $kn \times n$ matrix $W$, and is tight frame ($W^T W = I$).  Denote by ${Col}(W)$ the column space of $W$ (see Figure \ref{fig:proj_y}). The column space forms an $n$-dimensional subspace embedded in $\mathbb{R}^{kn}$. 
Since $W$ is tight frame, it can easily be verified that the distance between two vectors in the transform domain that are in $Col(W)$ is identical to their distance in the spatial domain; i.e., let $\z_1 = W \x_1$ and $\z_2 = W \x_2$. The vectors $\z_1$ and $\z_2$ are in $Col(W)$; thus,
\begin{equation}
	\|\x_1 - \x_2\| = \|\z_1 - \z_2\| 
\end{equation}
On the other hand, if $\z_1$ and $\z_2$ are two vectors in the transform domain that are not in $Col(W)$, their distance in the spatial domain is identical to their distance in the transform domain after projecting onto $Col(W)$. Namely, if $\x_1 = W^T \z_1$ and $\x_2=W^T \z_2$ (i.e., $\z_1, \z_2$ are two vectors in the transform domain, and $\x_1, \x_2$ are their counterparts in the spatial domain), then 
\begin{equation}
\label{eq:mse_proj}
	\|\x_1 - \x_2\| = \|P(\z_1)-P(\z_2)\|
\end{equation}
where $P(\z)=W W^T \z$ is the projection of vector $\z$ onto  $Col(W)$.

Denote by $\| \n\| = \| \y-\x  \|$  and $\| \n_w\| = \| \y_w-\x_w  \|$ the RMSE between $\x$ and $\y$ in the spatial domain and between $\x_w$ and $\y_w$ in the transform domain, respectively, as defined in Section~\ref{sec:redundant}. Since both vectors, $\x_w$ and $\y_w$, are in $Col(W)$, the above relations give readily that $\|\n \|=\|\n_w\|$. This was also verified in Equation~\ref{eq:n_w} above. After applying the shrinkage functions $\tilde \y_w=\psi(\y_w)$, however, the signal $\tilde \y_w$ is not necessarily in $Col(W)$. 
Denote by $\y^*$ the optimal possible reconstruction result\footnote{The solution is biased as we apply scalar mapping functions whereas the optimal mapping function should be a scalar field \cite{raphan2008optimal}.} and its representation in the transform domain by $\y^*_w=W \y^*$. Clearly $\y^*_w \in Col(W)$ and accordingly, the optimal reconstructed RMSE is:
$$
\| \n^* \|= \| \y^* -\x \| = \|\y^*_w - \x_w  \|
$$ 
The shrinkage functions, however, provide $\tilde \y_w$, which deviates from  $\y^*_w$ by $\tilde \r_w$ (see Figure \ref{fig:proj_y}):
\begin{equation}
\tilde  \r_w = \tilde \y_w -\y^*_w 
\end{equation}

\begin{theorem}
\label{th:6}
For an over-complete transform with redundancy $k$, the expected RMSE is bounded from above by:
\begin{equation}
	\| \tilde \n_w^S \|_E \leq \| \n^* \|_E + \frac{1}{\sqrt{k}} \| \tilde \r_w \|_E
\end{equation}
\end{theorem}
\noindent
where $\Delta=\| \tilde \n_w^S \|_E$ is the resulting RMSE in the spatial domain. In other words, the larger the redundancy, the closer the resulting RMSE is to the optimal one and the convergence rate goes like $1/\sqrt{k}$. 
\\ \\
\noindent
{\bf Proof 6:}
Recall that
\begin{align*}
| \tilde \n_w^{\cal S} \|_E \doteq \| \tilde \y_w^S - \x\|_E = \| P(\tilde \y_w) - \x_w\|_E = \\
=\| P(\y^*_w + \tilde \r_w) - \x_w\|_E = \| \y^*_w + P(\tilde  \r_w) -\x_w \|_E
\end{align*}
where the second equality is due to Equation~\ref{eq:mse_proj} and the fourth equality is due to the fact that $\y^*_w \in Col(W)$.
Using the triangular inequality, we get:
$$
\| \y^*_w + P(\tilde  \r_w) -\x_w \| \leq \| \y^*_w -\x_w  \| + \| P(\tilde  \r_w) \| = \| \n^* \| + \| P(\tilde \r_w) \|
$$
Thus,
\begin{equation}
\label{eq:nws}
\| \tilde \n_w^{\cal S} \| \leq \| \n^* \| + \| P(\tilde  \r_w) \|
\end{equation}
\noindent
We follow the same argument that was used in Appendix~A where we showed that due to the stationarity of natural images, for any $W$ (any redundancy rate), we have:
$
\| \tilde \n_w  \|_E = \|  \tilde \n_{u}   \|_E
$
where  $\tilde \n_w = \tilde \y_w - \x_w$ and $\tilde \n_u = \tilde \y_u - \x_u$.
This argument also holds if we switch $\x$ with $\y^*$ providing:
$$
\left \| \tilde \r_w   \right \|_E =
\left \|  \tilde \r_{u}   \right \|_E
$$
where  $\tilde \r_w = \tilde \y_w - \y^*_w$ and $\tilde \r_u = \tilde \y_u - \y^*_u$. This gives that the expected value of $\| \tilde \r_w \|$ is the same for any $W$ of any redundancy rate and it equals $\| \tilde \r_{u} \|$. Thus, the vector $\tilde \r_w \in \mathbb{R}^{kn}$ can be seen as a random vector in $\mathbb{R}^{kn}$ whose expected length is constant for any $W$. Since $P(\r_w)$ is an orthogonal projection of a random vector from $kn$-dimensional space onto an $n$-dimensional space, the expected length of $P(\r_w)$ is:

\begin{equation}
\label{eq:r_proj}
\| P(\tilde \r_w) \|_E = \frac{1}{\sqrt{k}} \| \tilde \r_w \|_E
\end{equation}
Moreover, the Johnson-Lindenstrauss Lemma shows that $\| P(\tilde \r_w) \|$ is fairly tight concentrated around $\| P(\tilde \r_w) \|_E$ (see \cite{dasgupta2003elementary}). Combining Equations \ref{eq:r_proj} and \ref{eq:nws}, we obtain the relation given in Theorem~6. Q.E.D.

%% file: results.tex
\section{Results}
In the previous sections we presented three different optimization schemes and their relations. In this section we test the empirical behavior of these methods on real data. The images we used for our experiments can be seen in Fig.~\ref{fig:testtrainIm}.

To test the three methods, we used the optimization scheme suggested in \cite{hel2008discriminative}. There the authors suggest modeling the shrinkage functions using piecewise linear mappings:
$$
\psi_k(\y_k) = {\cal M}_k(\y_k; \p_k)
$$
where $\p_k$ is a parameter vector controlling the piecewise function. Since the shrinkage function ${\cal M}_k$ is linear with respect to the parameter vector $\p_k$, optimizing for $\p_k$ can be solved in a closed form solution using a set of noisy images along with their clean counterparts.  In contrast to the statistical
approaches, this technique does not require any estimation of the prior model or of the noise characteristics. The SFs are designed to
perform ``optimally'' with respect to the given examples, under the
assumption that they will perform equally well with similar new examples. Using Methods 1 and 2, the optimization is performed on each $\p_k$ independently; however, for Method 3, all $\p_k$ are optimized simultaneously. For more information about the optimization and the implementation, the reader is referred to \cite{hel2008discriminative}.

In all the experiments described below, we used the
undecimated windowed Discrete Cosine Transform (DCT) as the image transform. Since the DCT transform is unitary, the undecimated DCT is a tight frame. Because of the undecimated form, each wavelet band can be calculated using a single 2D convolution (with the corresponding DCT
basis as the convolution kernel). Additionally, the inverse transform can be applied by convolving the rectified coefficients with the kernels forming $B_k^T$, which are the reflected (180 degree rotation) DCT kernels. More details are given in \cite{hel2008discriminative}.

\subsection{Comparing all methods for a single noise level}
\begin{figure*}[tbh]
\centering
\includegraphics[height=0.35\textwidth]{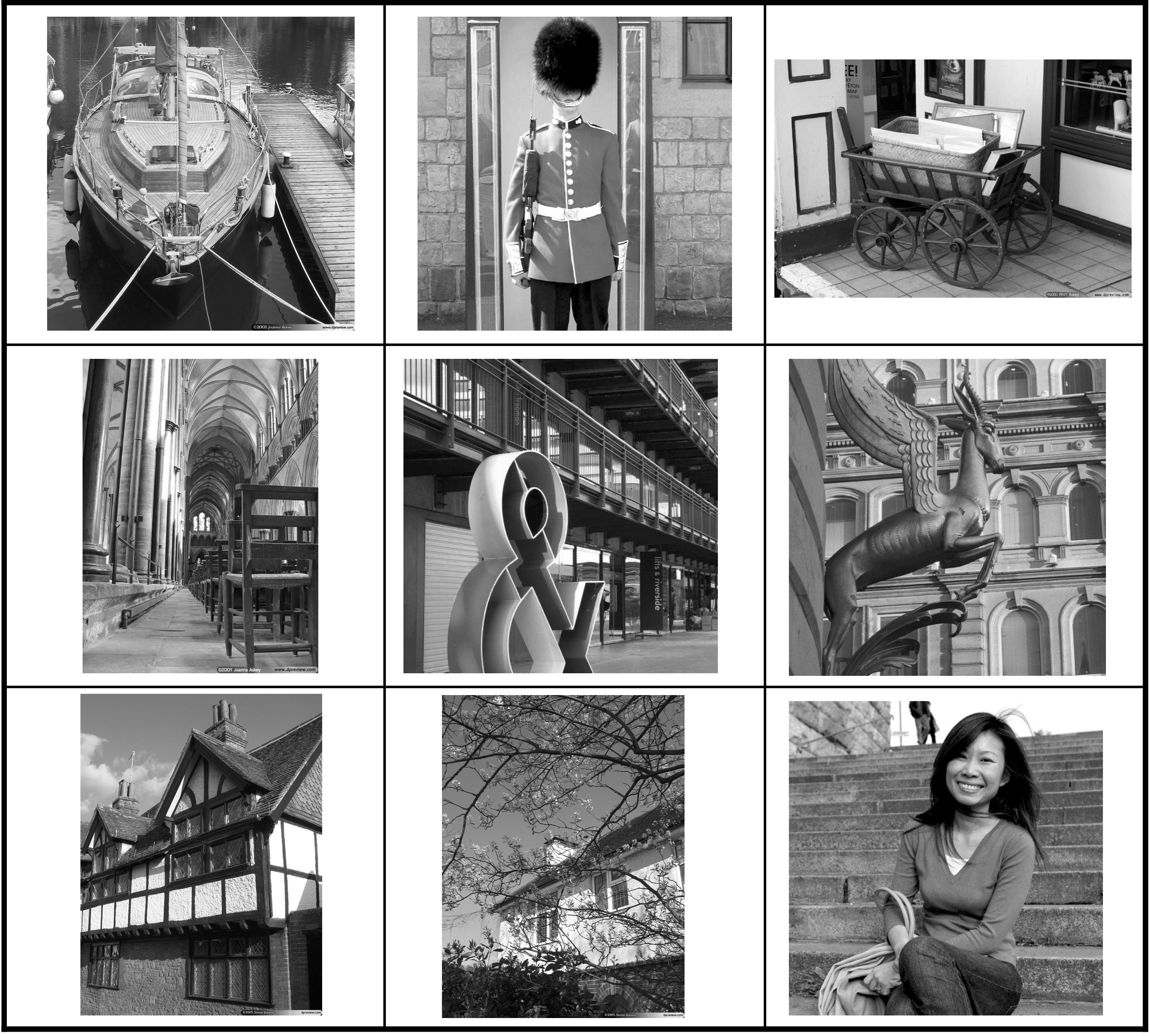}~~
\includegraphics[height=0.35\textwidth]{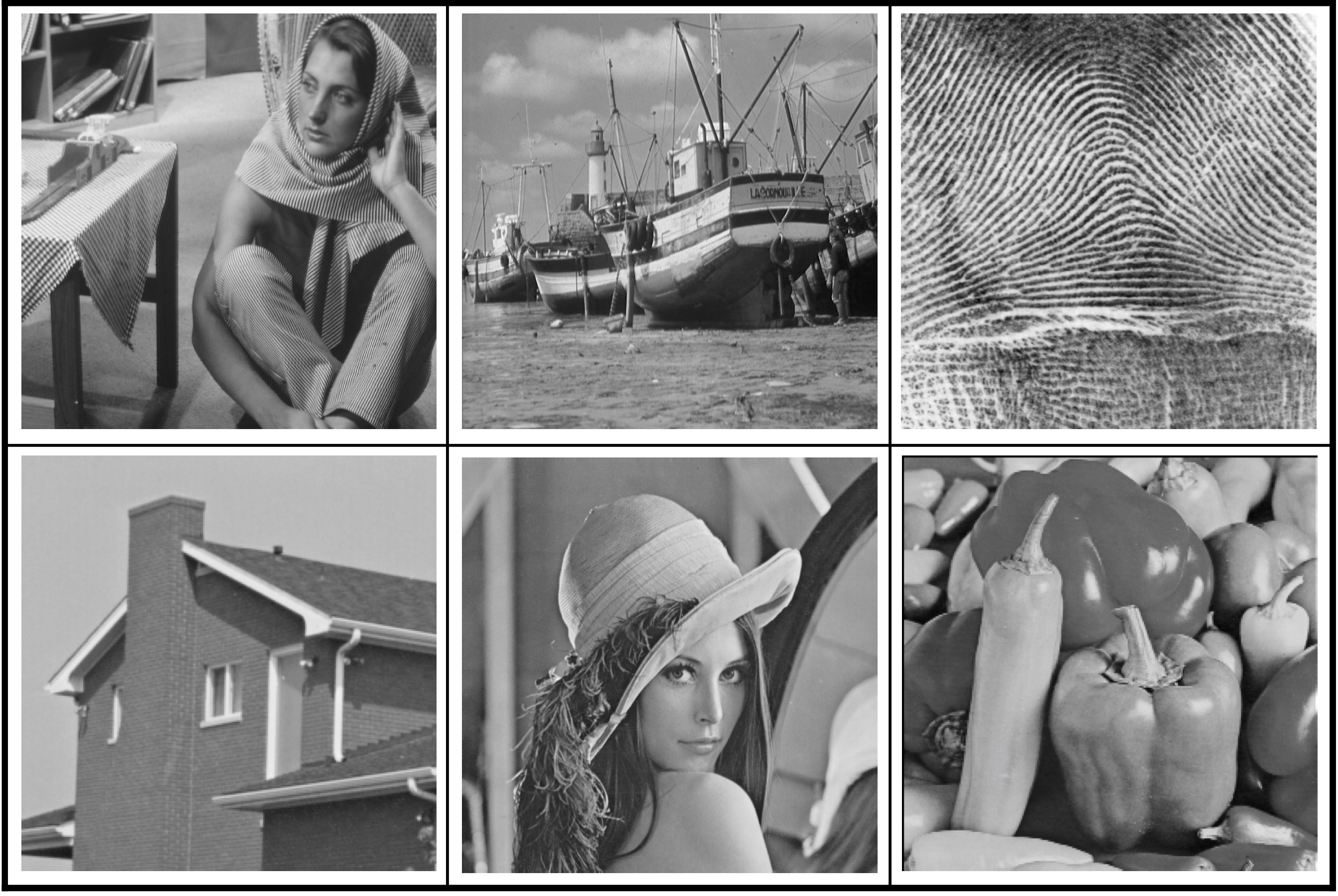}
\caption{
   Left: The images on which the SFs were trained.
Right: The images on which the denoising schemes were applied. 
} \label{fig:testtrainIm}
\end{figure*}

In the following experiments, unless mentioned otherwise, the
setting parameters were defined as follows: (1) Training images were grayscale natural images;  
a few of them are presented in Figure \ref{fig:testtrainIm}-left. (2) Test images were
taken from Figure \ref{fig:testtrainIm}-right. (3) Transform basis
was the undecimated $8 \times 8$ DCT. (4) The noise consists of
additive Gaussian noise with various STD values.

%\begin{figure}[tbh]
%\centering
%\includegraphics[width=0.4\textwidth]{Figures/trainIm.png}
%\caption{The images on which the SFs were trained.}
%\label{fig:trainIm}
%\end{figure}

\begin{figure*}[tbh]
  \centering 
  \includegraphics[width=0.8\textwidth]{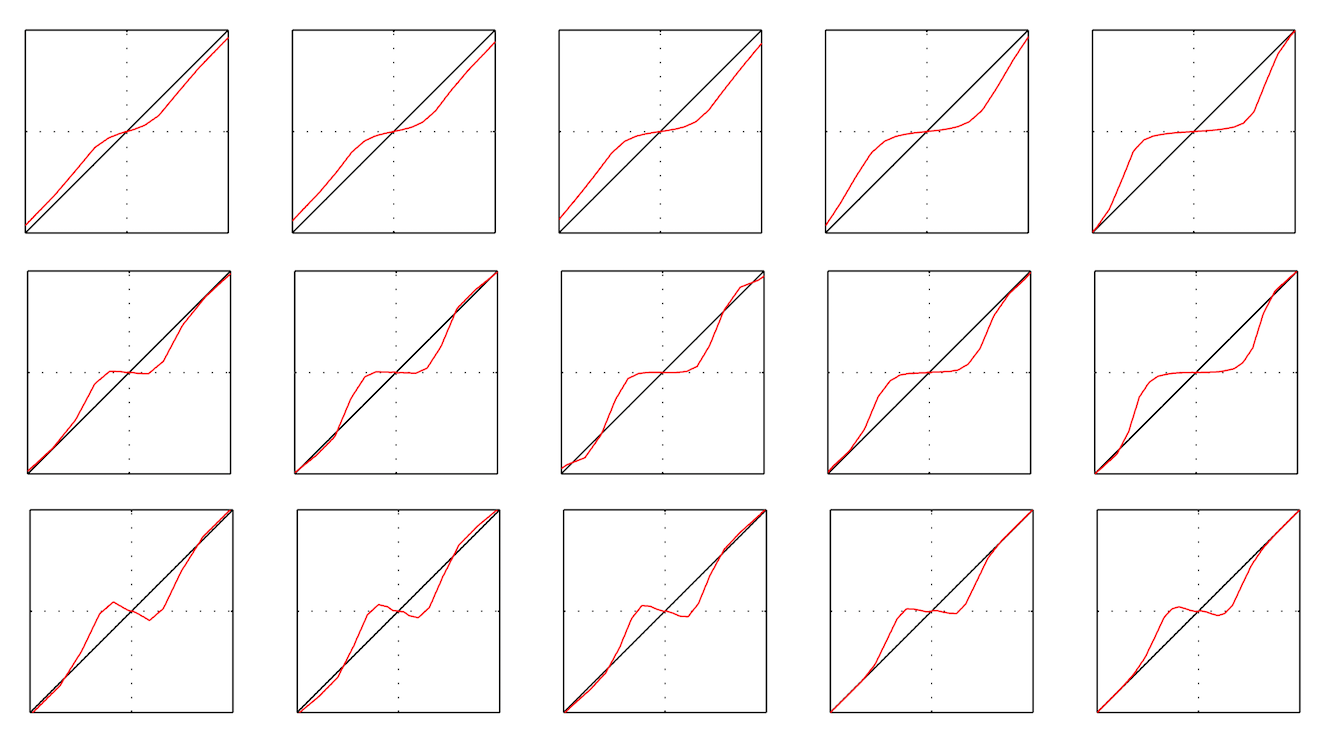}
    \parbox{0.95\textwidth} {  
    \caption{ Comparison of the produced SFs using Method 1 (top row), Method 2 (middle row),
   and Method 3 (bottom row).
   SFs on each row correspond to band $(i,i)$ of the $8 \times 8$ DCT basis,
   where  $i=2..6$ (left to right).
   Graph axes are shown in the range [-120,120]. 
\label{fig:SFm123}}}
\end{figure*}

Figure~\ref{fig:SFm123} displays some of the SFs
obtained for an $8 \times 8$ DCT basis, using the three methods described above for a noise level with a STD $\sigma =20$.
SFs on each row correspond to band indices $(i,i)$ of the $8 \times 8$ DCT basis,
where  $i=2..6$ (left to right).
Note that a DCT band with an index $(i,j)$ is the result of convolving
the image with a DCT basis
whose frequency is $i$ along the $x$-axis and $j$ along the
$y$-axis. 
The top, middle and bottom rows show the SFs
resulting from the first, second and the third methods, respectively. It can be seen that the SFs of the three methods are different from each other because each case takes into consideration different statistical correlations as explained above.

\begin{figure}[tbh]
  \centering
  \includegraphics[width=0.5\textwidth]{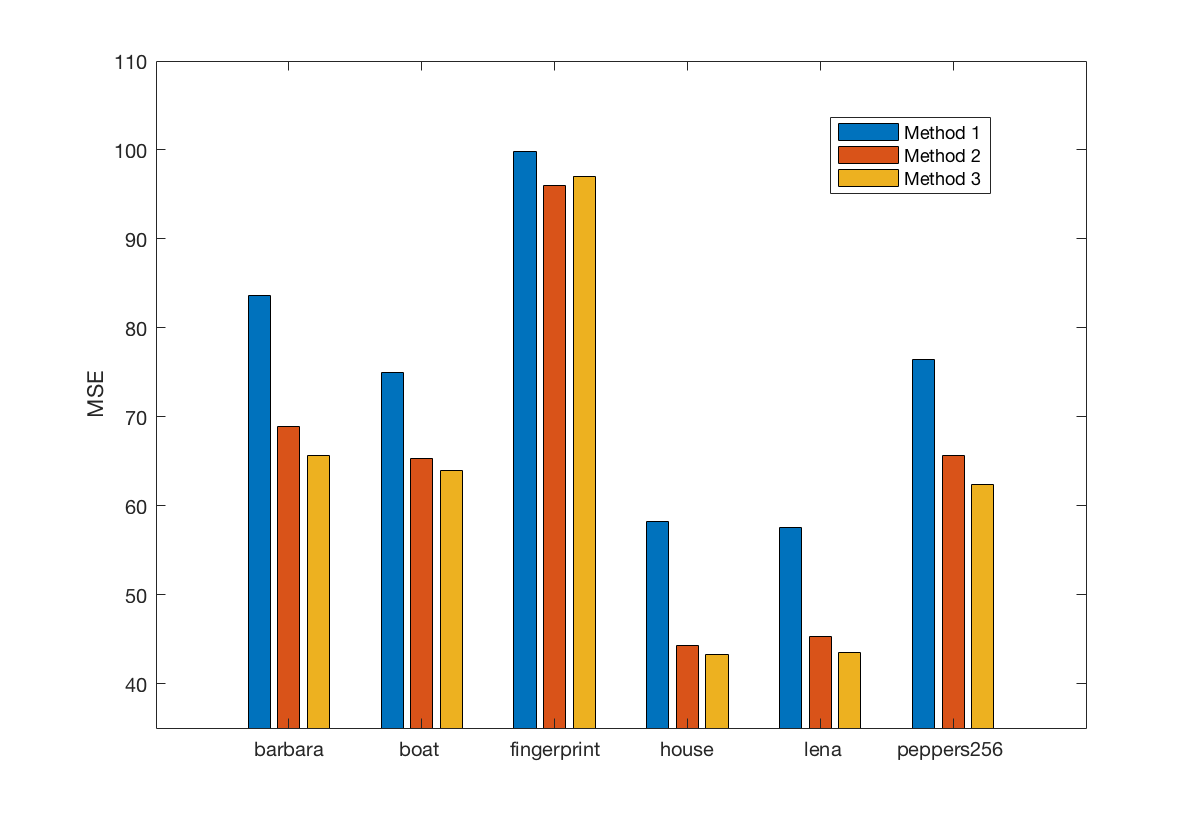}
\centering { \parbox{0.45\textwidth} {
\caption{
\label{fig2}
    MSE after applying the SFs produced by Methods 1--3. Each bar is an average over 10 different noise realizations.
    }
 }}
\end{figure}

The obtained SFs were applied to several images\footnote{Taken from
http://decsai.ugr.es/javier/denoise/test\_images/index.htm} shown in Figure \ref{fig:testtrainIm}-right.
Figure~\ref{fig2} compares the resulting MSE for each described method. It is composed of six clusters of bars,
each of which compares the denoising results of a particular image. Each bar presents the resulting MSE averaged over 10
realizations of noise with a STD $\sigma=20$. The results demonstrate the improvement of the second method over the first method,
and the superiority of the third method over the other two. It can be seen that most of
the improvement is achieved when applying the objective in the
spatial domain (Method 2). Further improvement, although less
significant, is achieved when incorporating the band dependencies (Method 3). 
Note, however, that the resulting MSE of the {\sc
fingerprint} image is better for Method 2 than for Method 3, and this result is incompatible with Equation~\ref{eq:c32}. The reason for this outcome is that the training set for this experiment does not seem to be a good representative of the statistics of the textured {\sc fingerprint} image. This means that $\hat \psi_2$, $\hat \psi_3$ are not necessarily the right SFs that minimize $\Delta_2$, $\Delta_3$, respectively. And, indeed, training the SFs from statisticaly similar images and applying them to the same noisy {\sc fingerprint} image, provides an MSE value of 89.55 for Method 3, compared to 97.07 in the current plot. 

\subsection{Dependencies on the training statistics}
\begin{figure*}[tbh]
\centering{
\includegraphics[width=0.45\textwidth]{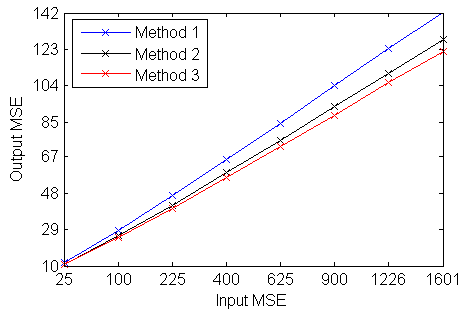}
\includegraphics[width=0.45\textwidth]{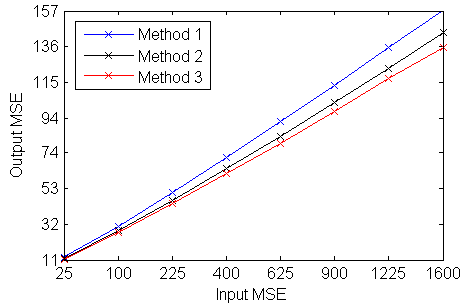}\\
~~~~~(a)~~~~~~~~~~~~~~~~~~~~~~~~~~~~~~~~~~~~~~~~~~~~~~~~~~~~~~~~~~~~~~~~~~~(b)\\
\caption{Comparison of the different optimization schemes for the case where the training and test images  (a) share the same statistics and (b) share different statistics. As expected, for both cases, the optimal optimization scheme is Method 3. The x-axis shows the input MSE and the y-axis presents the output MSE. For the first case, the input and output MSEs were computed from $30$ instances of test images, and for the second case, from $180$ instances.
\label{Fig:Bounds}}}
\end{figure*}

In the next experiments, we tested the relations between the different schemes for various noise levels and for two cases: (a) The training and test sets share the same statistics. (b) The training and test sets have different statistics. Performance was tested for eight different equally spaced noise levels, from STD $\sigma=5$ up to $\sigma=40$ . For the first case, we used the same image for training and test sets, where we applied five different noise realizations of the same noise level for each test. This was carried out for each image in the set. For the second case, we used a single image (in its turn) as the training set and the remaining images as the test set. The resulting MSEs for eight different noise levels are presented in Fig.~\ref{Fig:Bounds}. The x-axis shows the input MSE and the y-axis presents the output MSE. It can be seen that for both cases, $\Delta(\hat \psi_1) \geq \Delta (\hat \psi_2) \geq \Delta(\hat \psi_3)$. Thus, optimizing the SFs using Method 3 gives the optimal result while optimizing using Method 2 gives better bounds than the traditional optimization using Method~1.

\subsection {Cost-effective analysis}
\begin{figure*}[tbh]
\centering{
\includegraphics[width=0.42\textwidth]{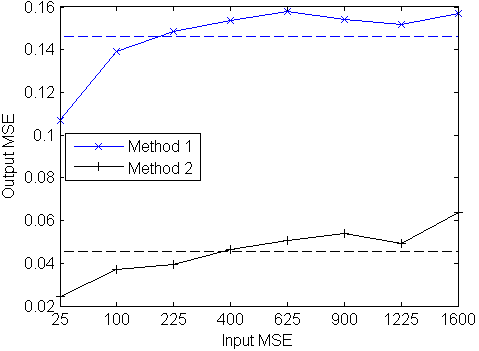}
\includegraphics[width=0.5\textwidth]{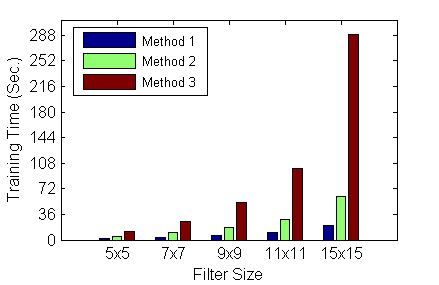}
\caption{
Left: The plot shows the relative deviation of the output MSE in Methods 1 and 2 relative to Method 3. The x-axis shows the input MSE and the y-axis presents the relative MSE with respect to Method~3. The deviation is approx. 4\% for Method 2 and approx. 16\% for Method 1.
Right: This plot shows the training time required for each method.
\label{Fig:CE1}
}}
\end{figure*}

 Optimizing SFs according to Method 2 offers both tight bounds on the mean MSE distortion and fast implementation. Figures ~\ref{Fig:CE1}-left shows the fraction of deviation of Methods 1 and 2 from the optimal scheme (Method 3). It can be seen that Method 2 deviates on average 4\% relative to Method~3, while Method 1 deviates about 16\%. 
 
 Figures ~\ref{Fig:CE1}-right shows the computation time taken to train the SFs for each method. It can be seen that training using the optimization scheme of Method 2 takes one quarter of the time needed by Method 3. Thus, Method 2 introduces a cost-efficient advantage; By allowing a deviation of 4\% from the optimal optimization scheme, we can get a gain, on average, an increase of 70\% in the speed. 

%\subsection{Single image training} Although not guaranteed theoretically, we can see that the above relations between the different optimization schemes are usually kept even if we the train the MFs on a single image. Figure \ref{Fig:NoisePerImage} shows the mean and s.t.d of each of the optimization for a set of trained images. Each bar cluster represents a training image where all the other images were the test images.
%
%\begin{figure}[htb]
%\centering{
%\includegraphics[width=0.49\textwidth]{ResultsFigures/noiseBarPerImageMean.png}
%\includegraphics[width=0.48\textwidth]{ResultsFigures/noiseBarPerImageSTD.png}\\
%(a)~~~~~~~~~~~~~~~~~~~~~~~~~~~~~~~~~~~~~~~~~~~~~~~~~~~~~~~~~~~~~~(b)
%
%\caption{The relations between the different optimization schemes are kept even if we train on a single image. Each bar represents a training image where all the other images were the test images. (a) The mean of the MSE distortion after the denoising. (b) The s.t.d of the MSE distortion after the denoising.
%\label{Fig:NoisePerImage}}}
%\end{figure}
%

\subsection{Denoising improvements vs. redundancy rate}
\begin{figure*}[tbh]
\centering{
\includegraphics[width=0.49\textwidth]{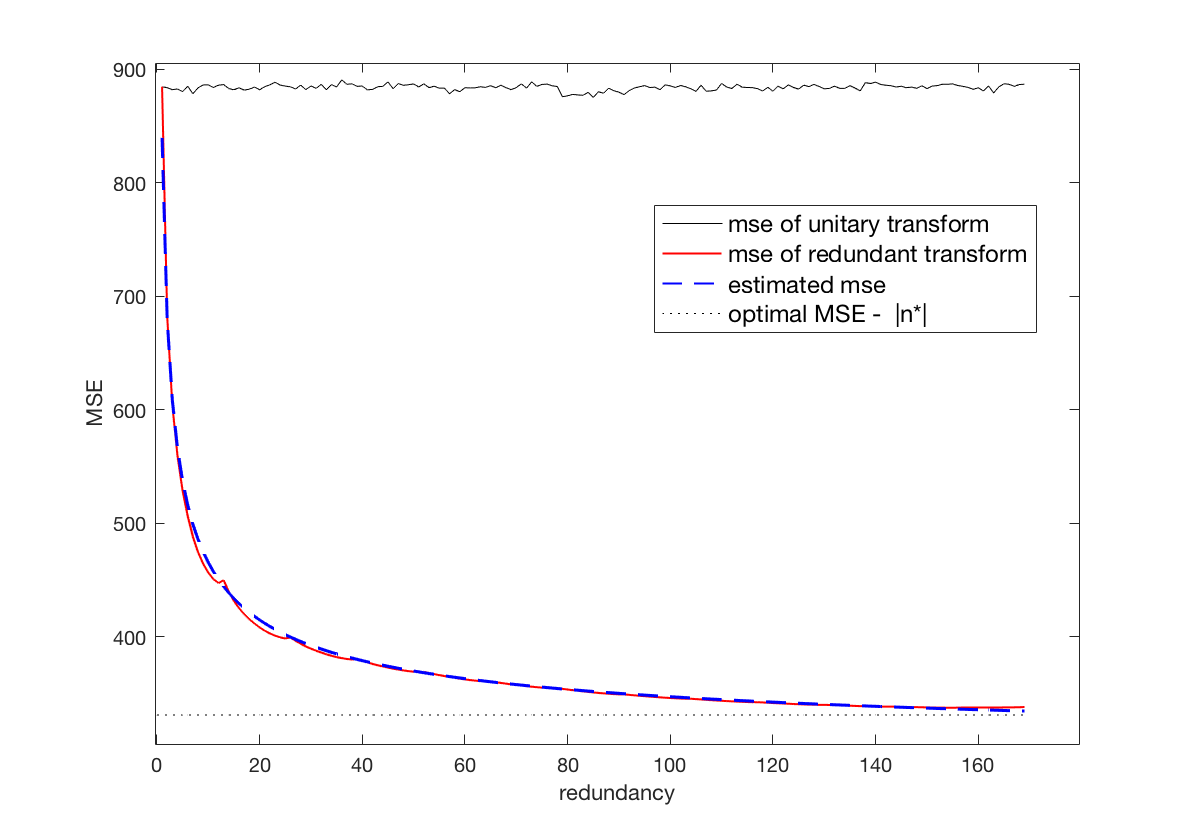}
\includegraphics[width=0.47\textwidth]{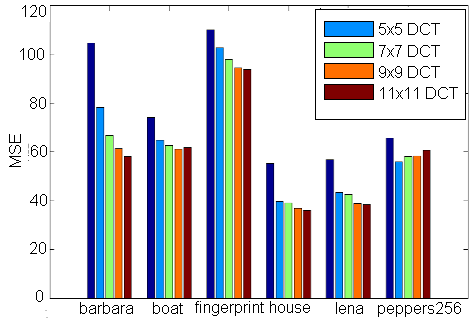} 
\caption{
\label{fig:Redundancy}
The performance is improved with the redundancy; The output MSE as a function of the redundancy rate. Left: The filter size was $13 \times 13$ and the noise STD was $\sigma=50$. The plot shows the RMSE as a function of redundancy rate. 
Right: The wider the filter, the more redundancy achieved and the better the denoising results.
}}
\end{figure*}

To validate the observation about the denoising improvement with respect to the redundancy rate, we measured the resulting MSEs for various redundancy rates. Figure~\ref{fig:Redundancy}-left shows the MSE for denoising applied to a $13 \times 13$ windowed DCT transform, where the redundancy rates were implemented by shifting the basis functions by $(i,j)$ along the x-axis and the y-axis, where $(i,j)\in \{0,12\} \times \{0,12\}$. Thus, the redundancy rate can range between $k=1$ and $k=169=13\times 13$. We assume the optimal MSE is given for $k=169$ (maximum redundancy) where in this case ${\hat \y} =\y^*$ and $\|\n^*\| = \|\x - \y^*\| = \mathcal{E}_{opt}$.
On the other hand, when $k=1$, we have that $\tilde \y_w \in Col(W)$  (in this case $W=U$) and thus $\|P(\tilde \r_w)\| = \|\tilde \r_u\| = \|  \tilde \y_u^S - \y^* \| = \Delta^*$. Accordingly, following Theorem~6, the RMSE should be:
\begin{equation}
\label{fig:mse_redund}
RMSE(k) \leq  \mathcal{E}_{opt} +  \frac{1}{\sqrt{k}} \Delta^*
\end{equation}

Figure \ref{fig:Redundancy}-left shows the decrease in the measured MSE as a function of the redundancy rate. The solid red curve shows the actual measures while the dashed blue curve shows the expected RMSE following Equation~\ref{fig:mse_redund}. It can be seen that the two plots basically overlap.
The measures were taken for the {\sc barbara} image where the noise level was $\sigma=50$. Each MSE measure in this plot is an average of five noise realizations. 

Figure \ref{fig:Redundancy}-right shows the optimal achieved MSE for each DCT transform, which were $5, 7, 9$ and $11$ pixels wide. It can be shown that, in most cases, the wider the filter, the more redundancy can achieved and the better the denoising results.

%\begin{figure}[tbh]
%\centering{
%\includegraphics[width=0.6\textwidth]{Figures/quality.png}
% \centering {\parbox{0.9\textwidth}{
% \caption{
% \label{fig:quality}
% 	The inherent tradeoff between Quality vs. Efficiency with respect to the three optimization methods.}
%  }}}
%\end{figure}
%

%To see this remember that $\Delta_3=\sum_i
%\Delta_2^i$ where $i$ runs over the various bands. This can be
%interpreted as a projection of the vector $\Delta_2$ onto a vector
%of ones denoted by $\bf 1$. Thus,
%$$
% \Delta_3(\p_2) = {\bf 1}^T \Delta_2(\p_2) ~~\mbox{and}~~
% \Delta_3(\p_1) = {\bf 1}^T \Delta_2(\p_1)
%$$
%Although the magnitudes $\| \Delta_2(\p_2) \|^2 \leq \|
%\Delta_2(\p_1) \|^2$, the projections onto the vector $\bf 1$ does
%not necessarily preserves the order of magnitudes. Nevertheless,
%this is true on the average case if we assume that $\Delta_2(\p_2)$
%and $\Delta_2(\p_1)$ are randomly distributed.
%

\section{Conclusions}
Transform denoising using shrinkage functions is a classical framework that is widely used in numerous applications. In this paper we provide a complete picture of the interrelations between the transform used, the optimal shrinkage functions, and the domains in which they are optimized. In particular, we provide a theoretical justification for applying the shrinkage functions in the transform domain and the benefit of applying them in redundant bases. 

Additionally, we provided theoretical bounds for the three possible optimization schemes of the shrinkage functions. We showed that for subband optimization, where each $\psi_i$ is optimized independently, optimizing the shrinkage function in the spatial domain is always better than or equal to optimizing them in the transform domain. This option, besides being simple to implement, is proven to outperform the traditional transform domain optimization while avoiding the demanding spatial domain optimization of all the shrinkage functions, simultaneously. 

For redundant bases, we provided the expected denoising gain we may achieve, relative to the unitary basis, as a function of the basis redundancy. This result allows a user to make a clever decision about the redundancy used by taking into account the expected denoising gain and the computational time allocated for this process.

%% file: appndx.tex
\appendices

\section {}

\noindent
{\bf Theorem 1:}
After denoising, the expected MSE distortions in the transform domain are equal for the unitary and for the redundant transforms, i.e.:
$$
\left \| \tilde \n_u   \right \|_E = \left \| \tilde \n_w   \right \|_E
$$

\noindent
{\bf Proof 1:} The MSE value $\left \| \tilde \n_u   \right \|_E^2$ is defined as:
$$
\left \| \tilde \n_u   \right \|_E^2 = E \left  \{ \left \| \tilde \n_u   \right \|^2 \right \}  = \int \left  \| U \x - \psi \left \{ U \y \right \} \right  \|^2 P(\x,\y) d\x d\y
$$
where $P(\x,\y)$ denotes the probability distribution function of $(\x,\y)$.
By changing variables, the above expression can be rewritten as:
$$
\left \| \tilde \n_u   \right \|_E^2  = \int \left  \| U S_i \x - \psi \left \{ U S_i \y \right \} \right  \|^2 P(S_i \x,S_i \y) |\det(S_i^T) | \ d\x d\y
$$
where $S_i$ is a shift operator by the $i^{th}$ displacement.
Now, we exploit the stationary property of natural images. This property gives that for each $S_i$:
$$
P(S_i \x,S_i \y) = P(\x,\y)
$$
Additionally, we can apply the (adjoint) shift operator to the transform basis, rather than to the images. Using the notation $U S_i = U_i$ (Equation \ref{eq:wi}) and having $|\det(S_i^T)|=1$, we get:
\begin{eqnarray}
\left \| \tilde \n_u   \right \|_E^2  = \int \left  \| U_i \x - \psi \left \{ U_i \y \right \} \right  \|^2 P(\x, \y) d\x d\y = \|  \tilde \n_{u_i}   \|_E^2
\end{eqnarray}

Now, since $\left \| \tilde \n_w   \right \|_E^2  = \frac{1}{n} \sum_{i=1}^n  \|  \tilde \n_{u_i}   \|_E^2 $, we conclude:
$$
\left \| \tilde \n_w   \right \|_E^2 = \frac{1}{n}  \sum_{i=1}^n
\left \|  \tilde \n_{u_i}   \right \|_E^2 =
\frac{1}{n}  \sum_{i=1}^n \left \|  \tilde \n_{u}   \right \|_E^2  =
\left \|  \tilde \n_{u}   \right \|_E^2 ~~~~\mbox{Q.E.D}.
$$

\section{}
\noindent
{\bf Theorem 2:}
For any given $\psi$,
$$
 \| \tilde \n_u^{\cal S}  \|_E  \geq
 \| \tilde \n_w^{\cal S}  \|_E
$$

\noindent
{\bf Proof 2:}  After Equation~\ref{eq:wtd} and Theorem~1, we have:
$$
\| \tilde \n_u^{\cal S}  \|_E =  \| \tilde \n_u   \|_E = \| \tilde \n_w   \|_E
$$
However, from Equation~\ref{eq:utd} it follows that
$$
\| \tilde \n_w   \|_E \geq \| \tilde \n_w^{\cal S} \|_E
$$

and therefore:
$$
\| \tilde \n_u^{\cal S}  \|_E  \geq \| \tilde \n_w^{\cal S} \|_E ~~~~~~\mbox{Q.E.D}.
$$